\documentclass[12pt,preprint]{aastex}
\usepackage{graphics,graphicx,rotating,amsmath}

\newcommand{\bbe}{{\boldsymbol \beta}}
\newcommand{\bth}{{\boldsymbol \theta}}
\newcommand{\bps}{{\boldsymbol \psi}}
\newcommand{\bde}{{\boldsymbol \delta}}
\newcommand{\bga}{{\boldsymbol \gamma}}
\newcommand{\bep}{{\boldsymbol \epsilon}}

\newcommand{\mbg}{{\mbox{\boldmath$g$}}}
\newcommand{\mbk}{{\mbox{\boldmath$k$}}}
\newcommand{\Real}[1]{{\rm Re}\left[ #1 \right]}
\newcommand{\rSN}{{\rm SN}}

\newcommand{\cZ}{{\cal Z}}

\newcommand{\lr}[1]{\left( #1 \right)}
\newcommand{\lrs}[1]{\left[ #1 \right]}

\newcommand{\IZt}{{I^{0}(\bbe)}}
\newcommand{\ILnsd}{{I^{Lnsd}}}
\newcommand{\ILnsdt}{{I^{Lnsd}(\bth)}}
\newcommand{\ISmd}{{I^{Smd}}}
\newcommand{\ISmdt}{{I^{Smd}(\bth)}}

\newcommand{\IZSmdt}{{I^{0Smd}(\bbe)}}
\newcommand{\IESmd}{{I^{ESmd}}}
\newcommand{\IESmdt}{{I^{ESmd}(\bth)}}
\newcommand{\IDec}{{I^{Dec}}}

\newcommand{\hIZk}{{\hat I^{0}(\mbk)}}

\newcommand{\hILnsdk}{{\hat I^{Lnsd}(\mbk)}}

\newcommand{\hISmdk}{{\hat I^{Smd}(\mbk)}}

\newcommand{\hIDeck}{{\hat I^{Dec}(\mbk)}}

\newcommand{\hIESmdk}{{\hat I^{ESmd}(\mbk)}}

\newcommand{\hIZSmdk}{{\hat I^{0Smd}(\mbk)}}

\begin{document}
\title{A New Method for PSF correction
using the Ellipticity of Re-smeared Artificial Images 
in Weak Gravitational Lensing Shear Analysis
}
\author{Yuki Okura\altaffilmark{1}} 
\email{yuki.okura@nao.ac.jp}

\author{Toshifumi Futamase\altaffilmark{2}}
\email{tof@astr.tohoku.ac.jp}

\altaffiltext{1}
 {National Astronomical Observatory of Japan, Tokyo 181-8588, Japan}
\altaffiltext{2}
 {Astronomical Institute, Tohoku University, Sendai 980-8578, Japan}

\begin{abstract}
We propose a new method for Point Spread Function (PSF) correction in weak gravitational lensing shear analysis
using an artificial image with the same ellipticity as the lensed image. 
This avoids the systematic error associated with the approximation in PSF correction used  in previous approaches.
We test the new method with simulated objects which have Gaussian or Cersic profiles smeared by a Gaussian PSF, 
and confirm that there is no systematic error. 
\end{abstract}

\section{Introduction}
Weak gravitational lensing has been widely recognized as a unique and very powerful method for studying not
only the mass distribution of the universe but also  cosmological parameters (Mellier 1999, Schneider 2006, Munshi et al. 2008). 
One of the most interesting aspects of this field is to accurately measure the cosmic shear which 
is coherent distortion of background galaxies induced by large scale structure of the universe, 
 because this effect depends on the evolution of the structure  which is influenced by the nature of  dark 
energy.   
There have been some detections of cosmic shear (Bacon et al 2000; Maoli et al 2001; Refregier et al 2002; Bacon et al 2003; Hamana et al 2003, Casertano et al 2003; van Waerbeke et al 2005; Massey et al 2005; Hoekstra et al 2006).
However, since the distortion is very weak and there are many sources of noise, 
 a very accurate measurement scheme is needed for the shapes of the huge number of background galaxies 
as well as a sophisticated analysis scheme for the measured shear which avoids systematic errors as much as possible.     
At the moment there are several plans for wide field surveys with a large enough number of galaxies 
to reduce the statistical error. This means the systematic errors in  shear  
analysis methods become larger than the statistical errors. In fact, even  current survey plans require 
systematic errors less than 1\% error (Hyper Suprime-Cam http://www.naoj.org/Projects/HSC/HSCProject.html, Dark Energy Survey http://www.darkenergysurvey.org/, Euclid http://sci.esa.int/euclid  etc.), 
and 0.1\% error for future planned surveys (lsst http://www.lsst.org).
There is another important point in the analysis for such  wide survey data.
For example, in the plan for the  HSC wide survey,  a few hundreds of millions of galaxies will be measured,
therefore slow methods are not realistic even if the methods have a high accuracy. 
Thus it is essential to develop a fast shear analysis scheme which is free from systematic errors.  

There have been many studies in this direction (Kaiser et al 1995, Bernstein \& Jarvis 2002; Refregier 2003; Kuijken et al. 2006; Miller et al. 2007; Kitching et al. 2008; Melchior 2011),  
which have been tested using simulated data (Heymans et al 2006, Massey et al 2007, Bridle et al 2010 and Kitching et al 2012). 
We also have developed new analysis methods, including the E-HOLICs method (E-HOLICs part1 Okura and Futamase 2011; part2 Okura and Futamase 2012; part3 Okura and Futamase 2013).
The E-HOLICs method avoids  systematic error caused by the approximation in the weight function 
by adopting an appropriate elliptical weight function for shape measurement. 
However, the E-HOLICs method uses an approximation for PSF correction similar to other approaches.
Therefore,  previous approaches including E-HOLICs, cannot correct the PSF effect in some conditions
(e.g. large PSF or high elliptical PSF).

In this paper we concentrate on PSF correction and propose a new method which is free from the 
approximations used in previous moments methods and thus free from systematic error associated with PSF correction.
The idea is to produce  an artificial image with the true ellipticity as the result of 
re-smearing the PSF smeared image of the lensed image with true ellipticity.
We refer to this as the Ellipticity of Re-smeared Artificial images method (ERA).

This paper is organized as follows:
In section 2, we explain our notation and definitions used in this paper.  
There we also describe the zero plane used in the ERA method. 
We then present PSF smearing and PSF correction as used in the ERA method in section 3. 
In section 4, we test this method using simple test images.
Finally we summarize the method and give some comments.

\section{Basics}
In this section we describe the notation, definitions and basic idea  used in this paper.
\subsection{Zero plane and zero image}
In this paper, we use the zero plane and zero image instead of the source plane. 
The idea of the zero plane is that the intrinsic ellipticity of the source comes from a (imaginary) circular source
(called the zero image) in the imaginary plane (the zero plane) (the details can be seen in E-HOLICs part 2).

Suppose we have the reduced shear of lensing and the intrinsic ellipticity respectively as $\mbg^L$ and $\mbg^I$,  then 
the relationship of the displacements between the zero plane $\tilde \bbe$, the source plane $\bbe$ and the lens plane $\bth$ 
are described as
\begin{eqnarray}
\tilde \bbe&=&\bbe -\mbg^I\bbe^*\\
\label{eq:Cshear}
\tilde \bbe&=&\bth -\mbg^C\bth^*,
\end{eqnarray}
where $\mbg^C$ is combined shear which is defined as 
\begin{eqnarray}
\mbg^C \equiv \frac{\mbg^I+\mbg^L}{1+\mbg^I\mbg^{L*}}.
\end{eqnarray}

This combined shear has information of the intrinsic ellipticity and the lensing reduced shear. 
Since the intrinsic ellipticities are random,  
the lensing reduced shear can be obtained by removing the information of intrinsic ellipticity as
\begin{eqnarray}
\label{eq:obtainLRS}
\left< \frac{\mbg^C-\mbg^L}{1-\mbg^C\mbg^{L*}} \right>=\left< \mbg^I \right>=0.
\end{eqnarray}

This shows that we can obtain the lensing reduced shear in two steps.
The first step is to obtain the combined shear from each object (Eq.\ref{eq:Cshear}) and the  
second step is to obtain the lensing reduced shear by  averaging (Eq.\ref{eq:obtainLRS}).

In this paper we consider only the first step,
so we consider only the relationship between the zero plane and the lens plane, 
and we use $\tilde \bbe$ as $\bbe$ and $\mbg^C$ as $\mbg$ for notational simplicity.

\subsection{Notation and Definitions}
From this point, we use the complex coordinate $\bth\equiv\theta_1+i\theta_2$ in the image plane
and $\bbe\equiv\beta_1+i\beta_2$ in the zero plane.
Reduced shear, which is defined by gravitational convergence $\kappa(\bth)$ and 
gravitational shear $\bga(\bth)$, is also complex, 
$\mbg(\bth)\equiv g_1(\bth)+ig_2(\bth)\equiv\bga(\bth)/(1-\kappa(\bth))$.
The coordinate is associated with each image and set out the coordinate origin at the centroid of the image,
which is defined by requiring that the dipole moment of the image vanishes.

We use $\cZ^N_M(I,\bep_W)$ as complex moments of image $I(\bth)$ measured with the weight function $W(\bth,\bep_W)$ 
which has an arbitrary profile and an ellipticity $\bep_W$ which is  defined as
\begin{eqnarray}
\cZ^N_M(I,\bep_W)&\equiv&\int d^2\theta \bth^N_M I(\bth) W(\bth,\bep_W)\\
\bth^N_M&\equiv&\bth^{\frac{N+M}{2}}\bth^{*\frac{N-M}{2}},
\end{eqnarray}
where $\bth^N_M$ is the higher order complex displacement from the centroid.
Then we simplify the notation of combinations of complex moments as
\begin{eqnarray}
\frac{\cZ^N_M(I,\bep)+\cZ^O_P(I',\bep)}{\cZ^Q_R(I,\bep)}=
\lrs{\frac{\cZ^N_M+\cZ^O_P(I',\bep)}{\cZ^Q_R}}_{(I,\bep)}
\end{eqnarray}

Although the profile of the weight function is arbitrary,
 it should be a realistic profile, and we use the elliptical Gaussian weight function in the following simulations.
If the weight function is elliptical Gaussian such as
\begin{eqnarray}
W(\bth,\bep_W)&=&{\rm exp}\lr{-\frac{\bth^2_0-\Real{\bep_W^*\bth^2_2}}{\sigma^2_W}},
\end{eqnarray}
the Gaussian scale $\sigma^2_W$ should be fixed as the condition that the signal to noise ratio (SN) of monopole moment 
of the image has a maximum.
In this paper, we refer to the scale of the weight function as the maximum SN scale,
and define the maximum SN radius $R_W$ as
\begin{eqnarray}
2R_W^2&\equiv&\sigma^2_W.
\end{eqnarray}

Here SN is defined as
\begin{eqnarray}
\rSN\equiv\frac{\int d^2\theta I(\bth)W(\bth,\bep_W)}{\sqrt{\int d^2\theta W^2(\bth,\bep_W)}}
=\frac{\cZ^0_0(\bth,\bep_W)}{\sqrt{\int d^2\theta W^2(\bth,\bep_W)}},
\end{eqnarray}
so the weight scale should be set as
\begin{eqnarray}
\sigma_W=\sqrt{2\lrs{\frac{\cZ^2_0-\Real{\bep_W^*\cZ^2_2}}{\cZ^0_0}}_{\lr{I,\bep_W}}}.
\end{eqnarray}
In the above, we require that the dipole moment vanishes, so
\begin{eqnarray}
\cZ^1_1(I,\bep_W)=0.
\end{eqnarray}

\subsection{The Relationship between Ellipticity and Reduced Shear}
The ellipticity of the image $I(\bth)$ is defined by the quadrupole moments as
\begin{eqnarray}
\bep\equiv\lrs{\frac{\cZ^2_2}{\cZ^2_0}}_{\lr{I,\bep}},
\end{eqnarray}
where the ellipticity of the weight function is set to the same value as the measured ellipticity. 
Displacements in the zero plane and the image plane are related as 
\begin{eqnarray}
\bbe=\lr{1-\kappa}\lr{\bth-\mbg\bth^*}.
\end{eqnarray}

The brightness distribution of the lensed image $\ILnsdt$ is made from the zero image $\IZt$ by lensing, 
so $\IZt=\ILnsdt$. The relationship between the ellipticity of the zero image $\bep^0$ and the lensed image $\bep^L$ is obtained as 
\begin{eqnarray}
\label{eq:Ede}
0=\bep^0=\frac{\bep^L-2\mbg+\mbg^2\bep^{L*}}{(1+g^2)-2\Real{\mbg^*\bep^L}},
\end{eqnarray}
where $g=|\mbg|$. 

Here, because $\mbg$ and $\bep^L$ have the same phase angle (so $\mbg^*\bep^L=\mbg\bep^{L*}$), we obtain 
\begin{eqnarray}
\label{eq:Ede2}
0=\bep^0=\frac{\bep^L-\bde}{1-\Real{\bde^*\bep^L}},
\end{eqnarray}
where $\bde\equiv 2\mbg/(1+g^2)$ is the complex distortion.
eq.\ref{eq:Ede2} can also be obtained by changing $\bep^0$ in eq.\ref{eq:Ede} to $(\bep^0-\mbg^2\bep^{0*})/(1-g^2)$.

Finally, we obtain
\begin{eqnarray}
\bep^L=\bde.
\end{eqnarray}
This result means that we obtain the complex distortion $\bde$ (and also reduced shear) from $\bep^L$ ellipticity 
of the lensed image $\ILnsd$

\section{PSF correction using the ERA method}
The observed image is not the lensed image but the result of various effects such as atmospheric turbulence, photon noise and 
pixelization of the lensed image. These various effects smear the image and change the ellipticity. The PSF is supposed to 
express the smearing. Thus we need to corrected the smearing effect to obtain the correct ellipticity. 
In this section, we explain the PSF effect and present PSF correction using the ERA method. 
\subsection{Point Spread Function Effect}
In  PSF correction, the smearing effect is supposed to be described as follows: 
\begin{eqnarray}
\label{eq:SMD}
\ISmdt=\int d^2\theta' \ILnsd(\bth-\bth')P(\bth'),
\end{eqnarray}

where $\ISmdt$ is the brightness distribution for the observed smeared image and $P(\bth)$ is the PSF. 
We denote the ellipticity of $\ISmdt$ as $\bep^{Smd}$. 
The PSF effect occurs not only for galaxies but also for stars. Since a star is a point source, 
the brightness distribution of the smeared image of a star gives the PSF at that position. 
Therefore, the  PSF in the field can be obtained by some sort of interpolation from their values at the positions of stars
(We do not consider this interpolation here). 

The PSF changes the ellipticity of the lensed image. For the measured moment, using a moment method such as KSB or E-HOLICs, 
one can calculate the following expression.  
\begin{eqnarray}
Z^N_M(\ISmd,0)&=&\int d^2\theta f(\bth)\ISmdt W(\bth)
\nonumber\\
&=&\int d^2\theta\int d^2\psi f(\bth+\bps)\ISmdt W(\bth+\bps)\ILnsdt q(\bps)
\nonumber\\
f(\bth+\bps)&=&\lr{\bth+\bps}^N_M
\end{eqnarray}
In the previous approach, PSF correction makes use of the following approximation:
\begin{eqnarray}
\label{eq:KSBAP}
f(\bth+\bps)&\approx&f(\bth)+f''(\bth)\bps^2.
\end{eqnarray}
However there is no guarantee that higher order terms are negligible and this causes systematic errors 
in the shear measurement (see Figures below). 

Another approach of PSF correction is to use  deconvolution in the following way. 
Eq.\ref{eq:SMD} is described in Fourier space as
\begin{eqnarray}
\hISmdk=\hILnsdk\hat P(\mbk),
\end{eqnarray}
(where quantities with a hat indicate Fourier transformed functions).

The lensed image is then obtained by deconvolution as
\begin{eqnarray}
\hILnsdk\approx\hIDeck=\frac{\hISmdk}{\hat P(\mbk)},
\end{eqnarray}
where $\hIDeck$ is the deconvolved image. 

In real analysis it sometimes happens that the denominator vanishes (referred to as the 0 dividing problem),
so the following modification is used. 
\begin{eqnarray}
\hILnsdk\approx\hIDeck=\hISmdk\frac{\hat P^*(\mbk)}{|\hat P(\mbk)|^2+C^{Dec}},
\end{eqnarray}
where $C^{Dec}$ is referred to as the deconvolution constant.

If a large value for the deconvolution constant is used, it becomes dominant,
so the deconvolved functions are far from the correct functions desired. 
Therefore the deconvolution constant should be taken as small as possible to avoid the 0 dividing problem. 

However, this approach has another problem. 
Pixel noise makes random count in the real/Fourier space,
and thus we need to adopt large deconvolution constants to avoid the problem in real analysis
which leads to an incorrect PSF correction.
Therefore,  the deconvolution method cannot be used in real analysis.

\subsection{New method to obtain the ellipticity of the lensed image}
\begin{figure*}[htbp]
\epsscale{0.5}
\plotone{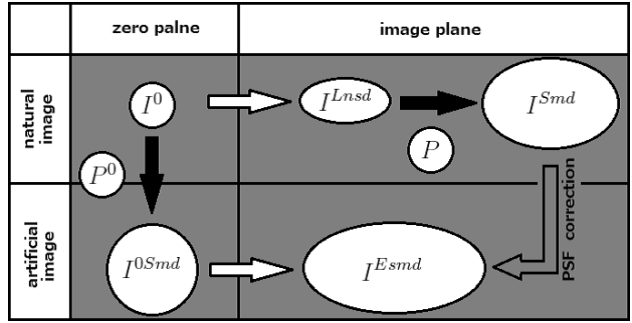}
\caption{
\label{fig:ERA_system1}
This figure shows the PSF correction using the ERA method.
White arrows indicate the lensing effect and black arrows indicate convolution.
}
\end{figure*}

The idea of this new PSF correction method is to produce an artificial image from the observed smeared image 
under the condition that  the artificial image  has the same ellipticity as the lensed image. 
Then we can obtain the ellipticity of the lensed image and weak lensing shear by measuring the ellipticity of 
the artificial image. This method is referred to as ERA (Ellipticity of Re-smeared Artificial image). 
In this section, we present the ERA method, and
Fig.\ref{fig:ERA_system1} shows a summary of the PSF correction.

Let's consider a circular image $\IZSmdt$ which is made from the zero image $\IZt$ by smearing with an arbitrary 
but circular PSF $P^0(\bbe)$ which is defined as
\begin{eqnarray}
\label{eq:ZsmdF}
\hIZSmdk\equiv\hIZk \hat P^0(\mbk).
\end{eqnarray}
and let's define $\IESmdt$ and $P^E(\bth)$ as brightness distributions made by the lensing effect from $\IZSmdt$ and $P^0(\bbe)$ respectively, so
\begin{eqnarray}
\IZSmdt&=&\IESmdt\\
P^0(\bbe)&=&P^E(\bth),
\end{eqnarray}

Because ellipticities of $\IZSmdt$ and $P^0(\bbe)$ are 0, their lensed image $\IESmdt$ and $P^E(\bth)$ 
have ellipticity $\bep^L$.
This means that smearing the lensed image $\ILnsdt$ by an artificial PSF $P^E(\bth)$ with ellipticity $\bep^L$ makes
artificial smeared image $\IESmdt$ with ellipticity $\bep^L$, and thus 
the PSF effect for measuring the ellipticity is corrected by transforming the real PSF $P(\bth)$ to $P^E(\bth)$ and 
making the artificial image $\IESmdt$ as follows
\begin{eqnarray}
\label{eq:ERA}
\hIESmdk=\hILnsdk \hat P^E(\mbk)=\hILnsdk \hat P(\mbk)\frac{\hat P^E(\mbk)}{\hat P(\mbk)}
=\hISmdk \frac{\hat P^E(\mbk)}{\hat P(\mbk)},
\end{eqnarray}
In this way, we obtain the PSF corrected ellipticity $\bep^L$ by   
measuring the ellipticity of $\IESmdt$ . 
This equation is satisfied only when the $\IESmdt$ and ${\hat P(\mbk)^E}$ have the same 
ellipticity $\bep^L$, given the observed ${\hat P(\mbk)}$ and $\hILnsdk$, and thus  
this equation can be solved somehow.

There are two possible ways to solve this equation.
One is to use  $P^E(\bth)$ directly, which means that the PSF correction is the re-smearing deconvolved image. 
However, because this method cannot avoid the problem of 0 dividing, 
eq.\ref{eq:ERA} must be modified as follows
\begin{eqnarray}
\label{eq:ERAREC}
\hIESmdk=\hISmdk \frac{\hat P^E(\mbk)\hat P^*(\mbk)}{|\hat P(\mbk)|^2+C^{Dec}}.
\end{eqnarray}
One solves this equation starting from the initial ellipticity $\bep^{Smd}$ under the condition 
that the ellipticities of $\IESmd$ and $P^E(\bth)$ have same values (referred to as method A). 

\begin{figure*}[htbp]
\epsscale{0.5}
\plotone{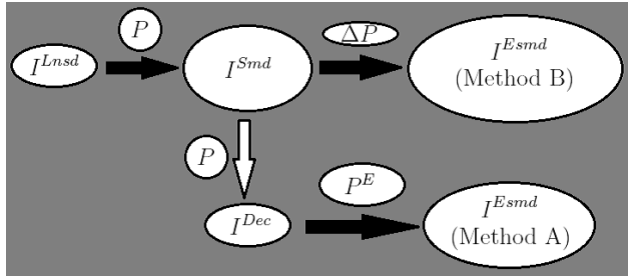}
\caption{
\label{fig:ERA_system2}
This figure shows the relationship between the images using the ERA method.
Black arrows mean convolution and white arrow means deconvolution by functions which are attached to the arrows.
$\IESmd$ has an arbitrary size and depends on the size of $\Delta P$ or $P^E$.
}
\end{figure*}

Another possibility is to introduce the correction for $PSF$ in the following way:
\begin{eqnarray}
\label{eq:ERAREP}
\hat P^E(\mbk)=\hat P(\mbk)\Delta \hat P(\mbk),
\end{eqnarray}
This cancels $\hat P(\mbk)$ in the denominator and avoids the problem of 0 dividing,
so Eq.\ref{eq:ERA} becomes
\begin{eqnarray}
\label{eq:ERARES}
\hIESmdk=\hISmdk \Delta\hat P(\mbk).
\end{eqnarray}
This method solves Eq.28 and Eq.29 simultaneously to find $\Delta\hat P^E$ under the condition that  
$\IESmd$ and $\hat P^E$ have the same ellipticity (referred to as method B).
Figure \ref{fig:ERA_system2} shows the relationship between the images.

Comparing these two methods,
Method A  has a deconvolution constant, so we must be careful about the effect from it,
and Method B creates and measures two images so
it is expected that a longer time is needed to obtain the PSF corrected ellipticity compared with method A.

\section{Simulation test}
In this section we perform several tests of the ERA method using simulated images and show the results. 
In the previous section, we introduced the artificial PSF $P^E(\bth)$ and the artificial image $I^{ESmd}$ 
with the lensed ellipticity,   
but did not mention their profile and size. In fact, these are both very flexible. 
In this test, we choose a Gaussian profile and use several sizes.  
A Gaussian profile is a standard choice and is appropriate for this initial test.
Investigations to identify a better profile and size of the artificial PSF will be part of future work to further develop the ERA method.
Parameters   used in this test are as described below.

\subsection{Simulation parameters}
We used the following parameters for the simulation test. 
The profile of the lensed image is assumed to be  [Gaussian, Cersic n=4] with  
an ellipticity $\bep^L=$(0.3,0.0) and
the size is set based on  the condition that the maximum SN radius is $R_W=20$ pixels.
This size is larger than the standard image used in real analysis,
because we would like to avoid the effect of pixelization in this test.
The ellipticity of the PSF $\bep^P$=[(0.0, 0.0), (0.1, 0.1), (0.3, 0.0), (-0.6, -0.6)] and 
PSF size ratio to Gaussian weight size of the lensed image is $R^P=$[0.5, 1.0, 1.5].

The first step of the simulation is to make images.
The lensed and PSF images are made with assigned parameters,
and we then make the smeared image $\ISmd$ in Fourier space from the lensed and PSF images.

PSF corrections  proceed as follows:
\begin{itemize}
\item Deconvolution: Deconvolve the smeared image $\ISmd$ by the PSF image $P$ for making deconvolved image $I^{Dec}$,
and measure ellipticity of the deconvolved image,
where we use a sufficiently small deconvolution constant ($C^{Dec} << |\hat P(\mbk)|^2$)
(In this test we do not have noise and thus we can take an arbitrarily small deconvolution constant).

\item Method A: Re-smear the deconvolved image $I^{Dec}$ by $P^E$ for making $\IESmd$,
and measure the ellipticity of $\IESmd$ which has same ellipticity as $P^E$.
To find the ellipticity of $P^E$, we use an iterative technique. 
The initial choice of the ellipticity for $P^E$ is that of $\ISmd$, and then smear $\IDec$ to 
make  $\IESmd$. Then, the measured ellipticity of the $\IESmd$ is used as the ellipticity of $P^E$ in the next iteration
(i.e. $\bep^{P^E}_{n}=\bep^{ESmd}_{n-1}$ in n-th iteration).

\item Method B: Re-smear the smeared image $\ISmd$ and PSF image $P$ by $\Delta P$ for making $\IESmd$ and $P^E$.
To find the ellipticity of $\Delta P$, we use the following iterative technique.
The initial choice of the ellipticity for $\Delta P$ is again that of $\ISmd$ to  make $\IESmd$ and $P^E$.
These ellipticities do not coincide in general, then 
the difference is added for the initial choice of the ellipticity for  $\Delta P$ in the second step and so on
(i.e. $\bep^{\Delta P}_{n}=\bep^{\Delta P}_{n-1}+(\bep^{ESmd}_{n-1}-\bep^{P^E}_{n-1})$ in n-th iteration).
\end{itemize}

As a comment on the iterative techniques used in Methods A and B,
we believe that the iterative methods used are standard,
but it is important to find a  faster and more robust method
which will be part of future studies.

Next, we test the following seven cases. 
\begin{itemize}
\item 1. Deconvolution: standard deconvolution where we use a small deconvolution constant, 
which  is not realistic due to pixel noise effect,
\item 2. Method A1 : size of $P^E$ is the same as the PSF $R^P$,
\item 3. Method A2 : size of $P^E$ is 2 times as large as  the PSF $R^P$,
\item 4. Method A3 : size of $P^E$ is 3 times as large as the PSF $R^P$,
\item 5. Method B1 : size of $\Delta P$ is the same  size as $R^P$, 
\item 6. Method B2 : size of $\Delta P$ is the same  size as $\ISmd$.
\item 7. KSB Method : PSF correction using the KSB method.
\end{itemize}

\subsection{Simulation results}
\begin{figure*}[htbp]
\epsscale{0.5}
\plotone{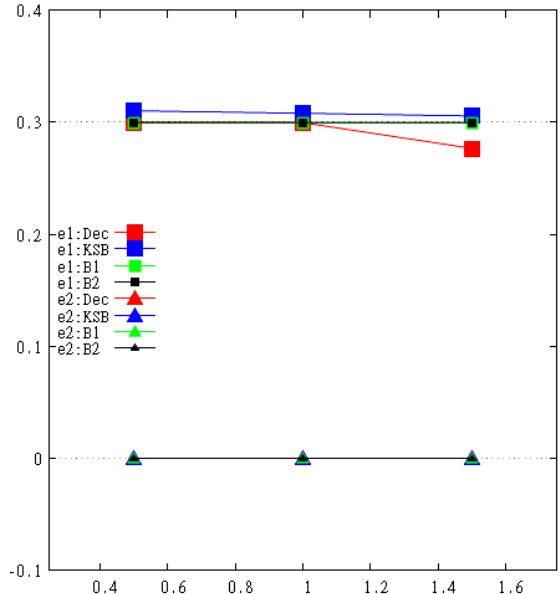}
\caption{
\label{fig:G_0_0A}
The results of the PSF correction test.
The PSF size ratio (0.5, 1.0, 1.5) is plotted on the horizontal axis and  ellipticity is on the vertical axis.
Square plots mean ellipticity1 with a true value of 0.3, and 
Triangle plots mean ellipticity2 with a true value of 0.0.
Red symbols (Dec) mean PSF correction by deconvolution,
Blue symbols (KSB) mean PSF correction using the KSB method,
green symbols (B1) indicate Method B1 and
black symbols (B2) indicate Method B2.
The profile of the lensed image is Gaussian,
and the ellipticity of the PSF is (0.0, 0.0).
}
\end{figure*}

\begin{figure*}[htbp]
\epsscale{0.5}
\plotone{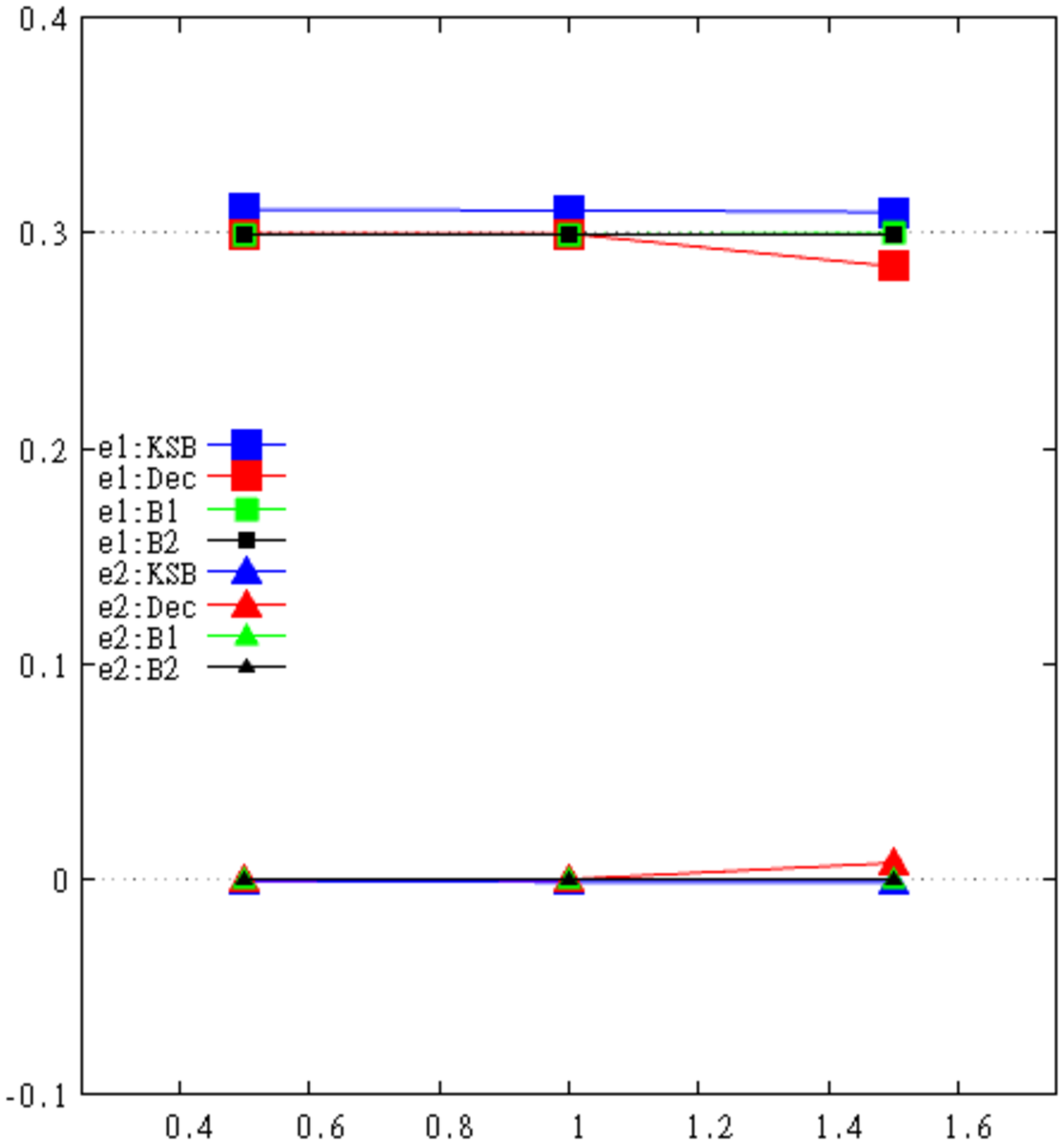}
\caption{
\label{fig:G_1_1A}
This figure is the same as Figure \ref{fig:G_0_0A} 
except the ellipticity of the PSF is (0.1, 0.1).
}
\end{figure*}

\begin{figure*}[htbp]
\epsscale{0.5}
\plotone{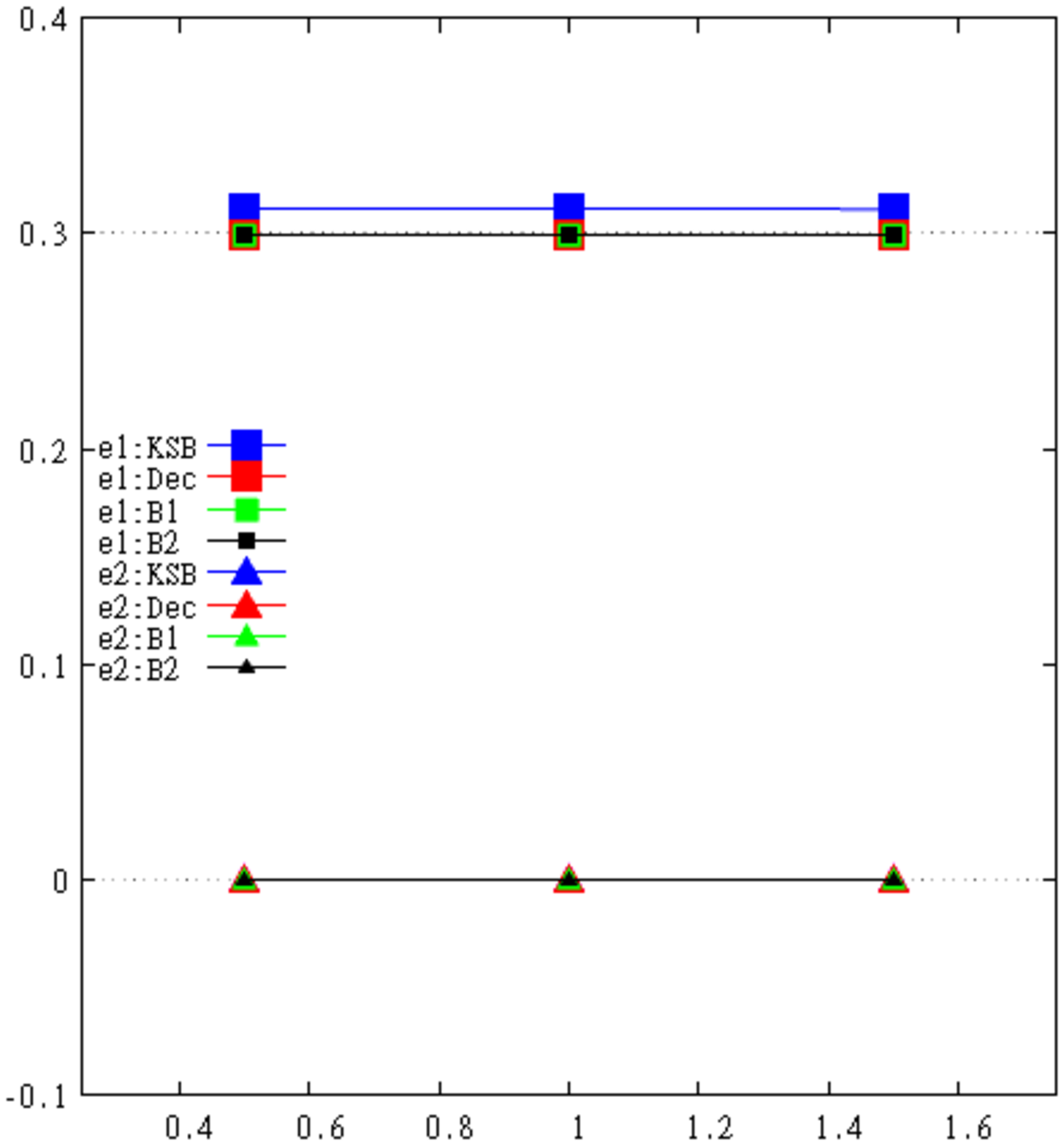}
\caption{
\label{fig:G_3_0A}
This figure is the same as Figure \ref{fig:G_0_0A} 
except the ellipticity of the PSF is (0.3, 0.0).
}
\end{figure*}

\begin{figure*}[htbp]
\epsscale{0.5}
\plotone{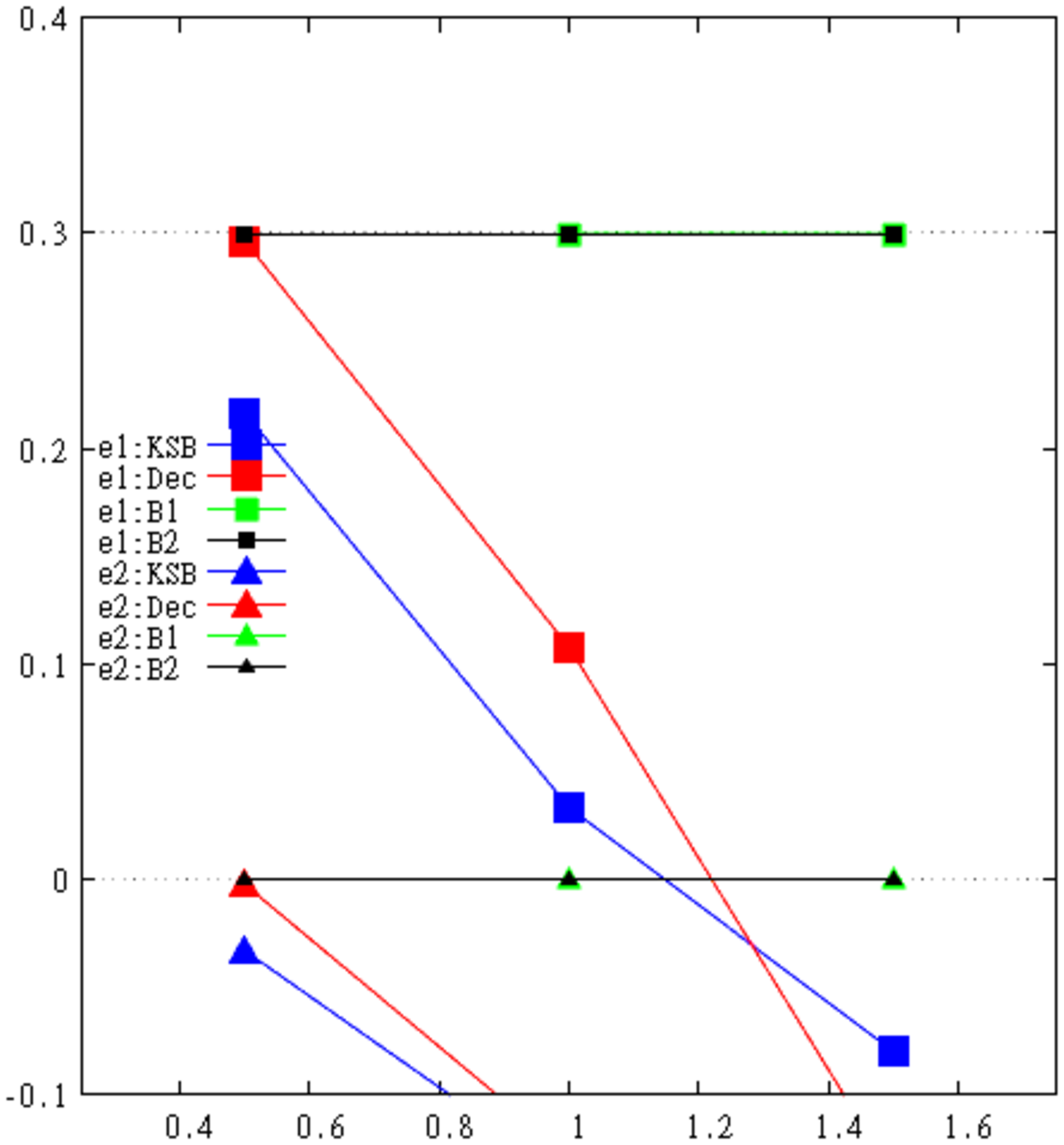}
\caption{
\label{fig:G_6_6A}
This figure is the same as Figure \ref{fig:G_0_0A} 
except the ellipticity of the PSF is (-0.6, -0.6).
}
\end{figure*}

\begin{figure*}[htbp]
\epsscale{0.5}
\plotone{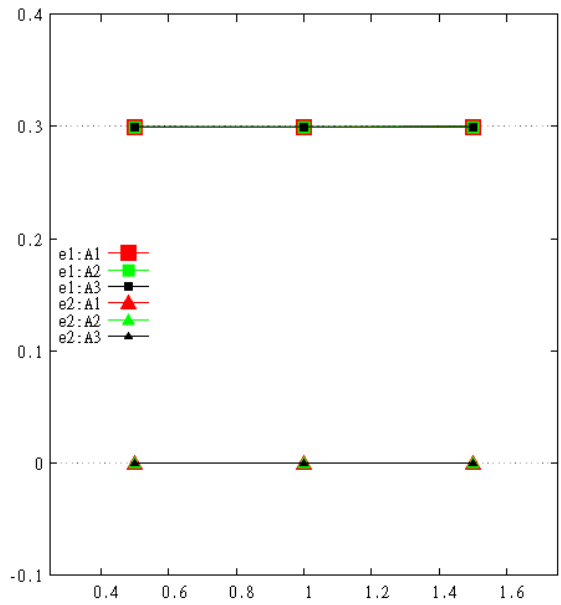}
\caption{
\label{fig:G_0_0B}
This figure is same as Figure \ref{fig:G_0_0A} 
except for the method used.
Red symbols (A1) mean PSF correction by Method A1,
green symbols (A2) indicate Method A2 and
black symbols (A3) indicate Method A3.
}
\end{figure*}

\begin{figure*}[htbp]
\epsscale{0.5}
\plotone{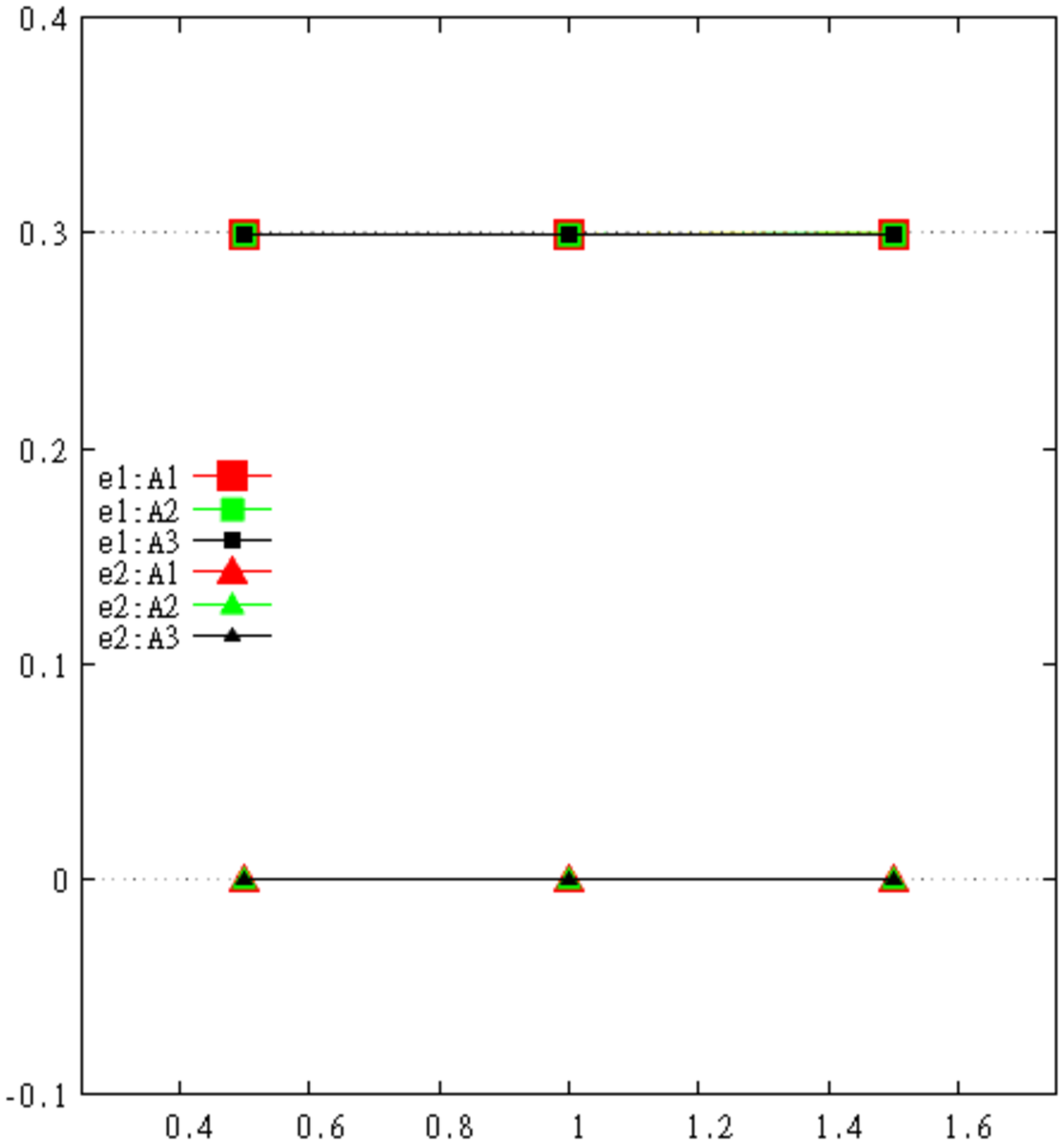}
\caption{
\label{fig:G_1_1B}
This figure is the same as Figure \ref{fig:G_0_0B} 
except the ellipticity of the PSF is (0.1, 0.1).
}
\end{figure*}

\begin{figure*}[htbp]
\epsscale{0.5}
\plotone{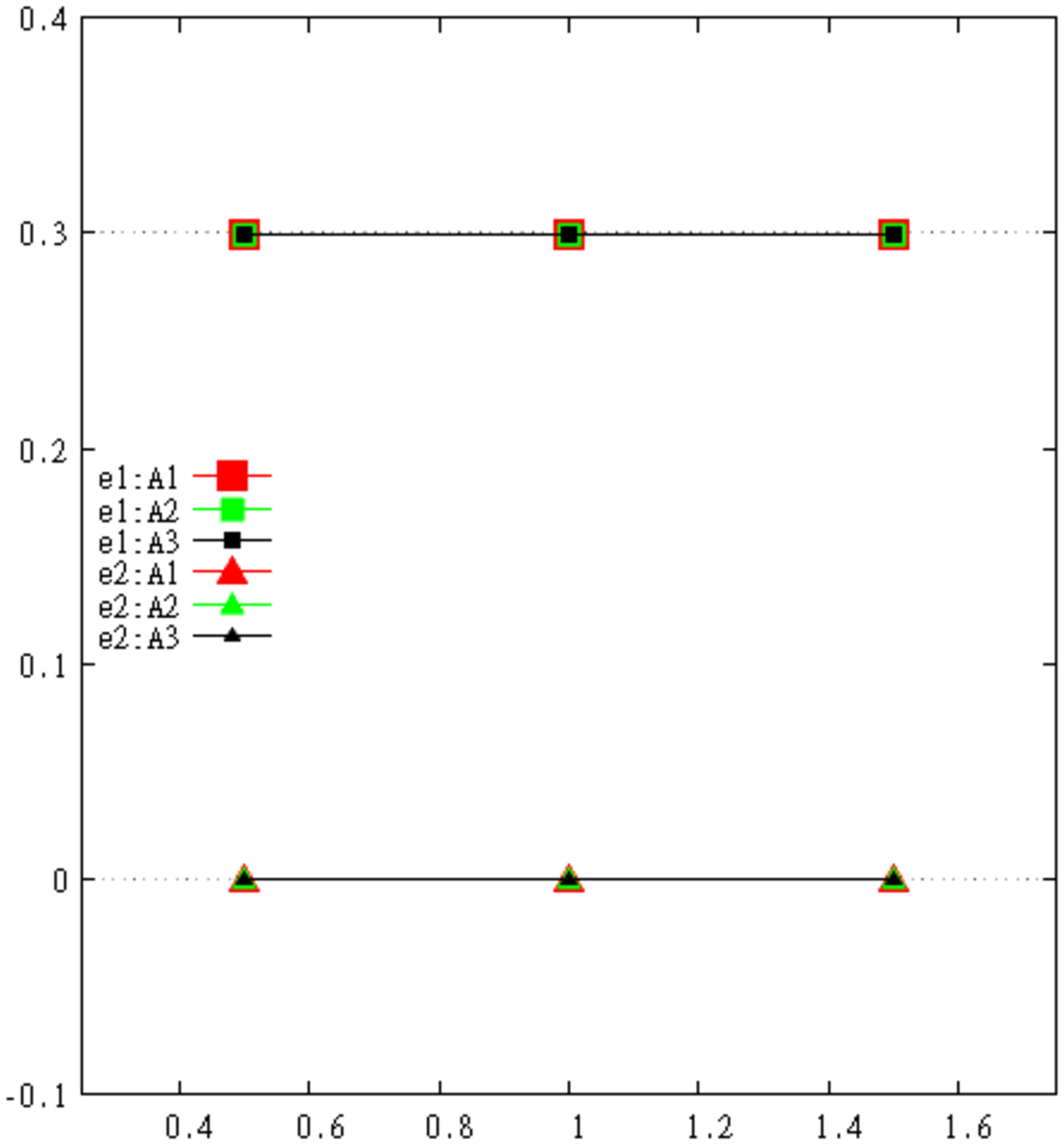}
\caption{
\label{fig:G_3_0B}
This figure is the same as Figure \ref{fig:G_0_0B} 
except the ellipticity of the PSF is (0.3, 0.0).
}
\end{figure*}

\begin{figure*}[htbp]
\epsscale{0.5}
\plotone{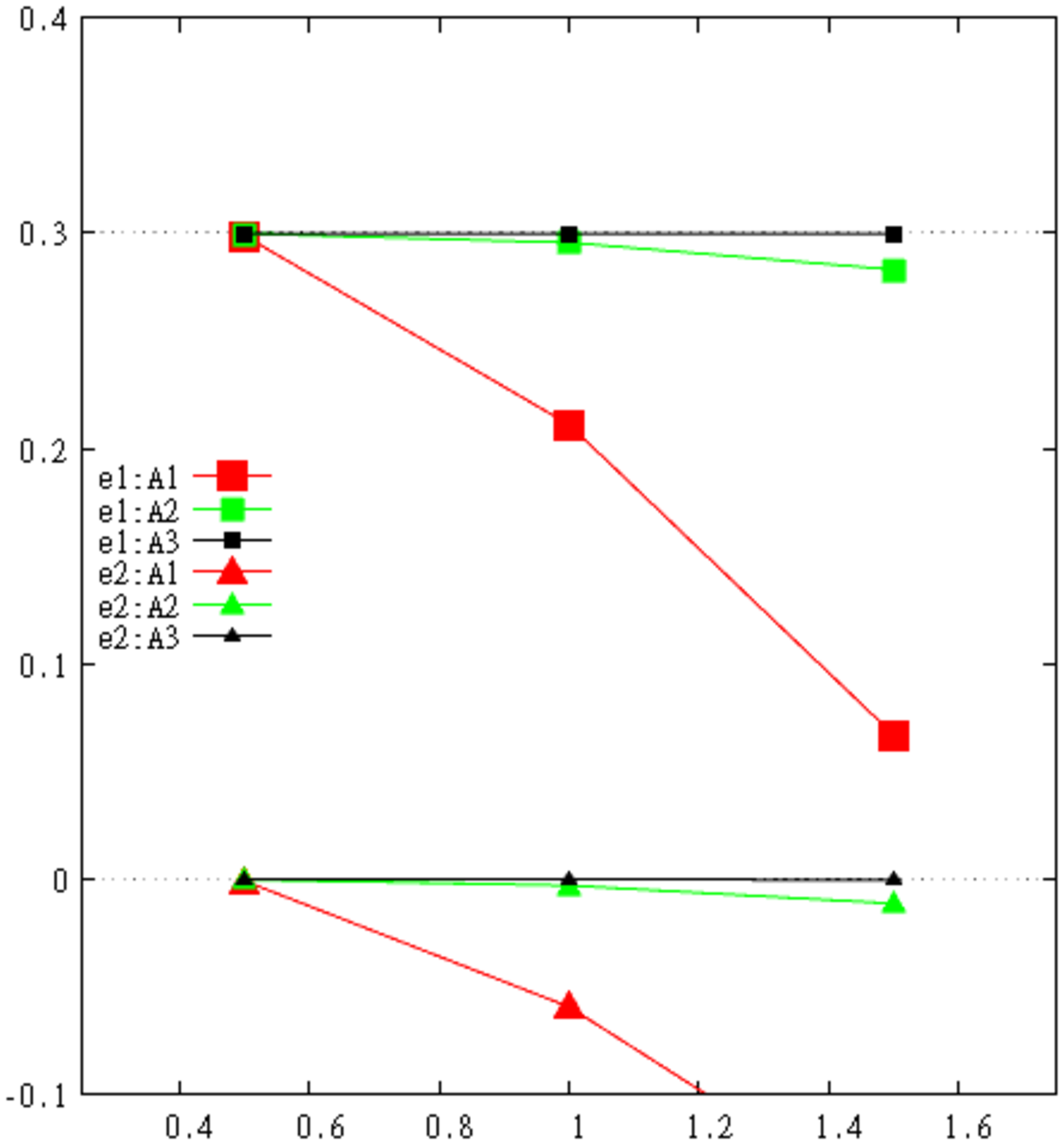}
\caption{
\label{fig:G_6_6B}
This figure is the same as Figure \ref{fig:G_0_0B} 
except the ellipticity of the PSF is (-0.6, -0.6).
}
\end{figure*}

\begin{figure*}[htbp]
\epsscale{0.5}
\plotone{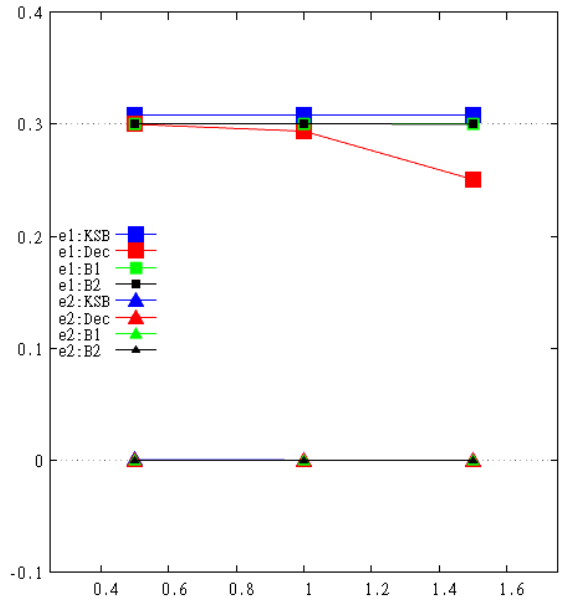}
\caption{
\label{fig:S_0_0A}
This figure is the same as Figure \ref{fig:G_0_0A} 
except the profile of the lensed image is Cersic.
}
\end{figure*}

\begin{figure*}[htbp]
\epsscale{0.5}
\plotone{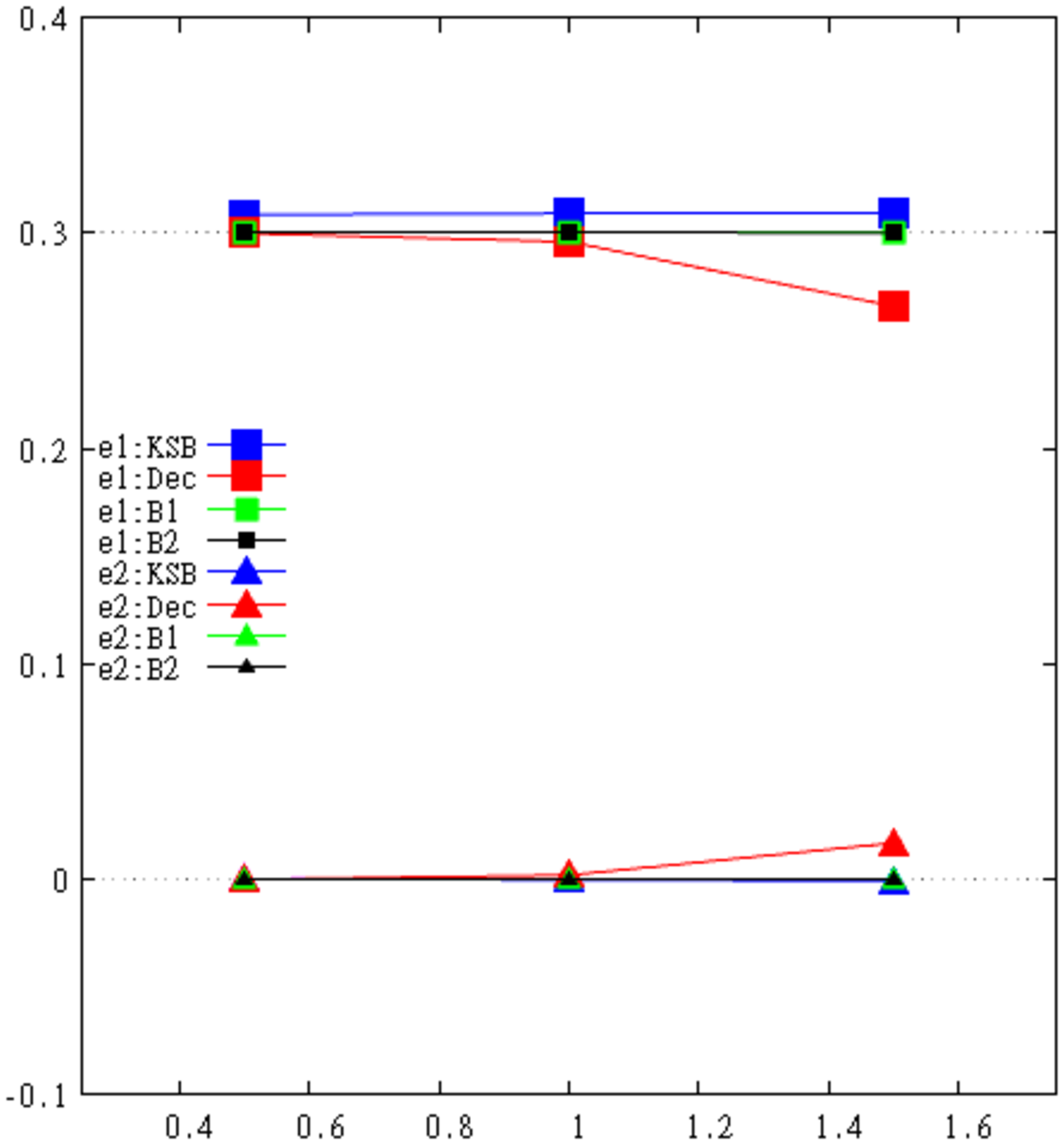}
\caption{
\label{fig:S_1_1A}
This figure is the same as Figure \ref{fig:S_0_0A} 
except the ellipticity of the PSF is (0.1, 0.1).
}
\end{figure*}

\begin{figure*}[htbp]
\epsscale{0.5}
\plotone{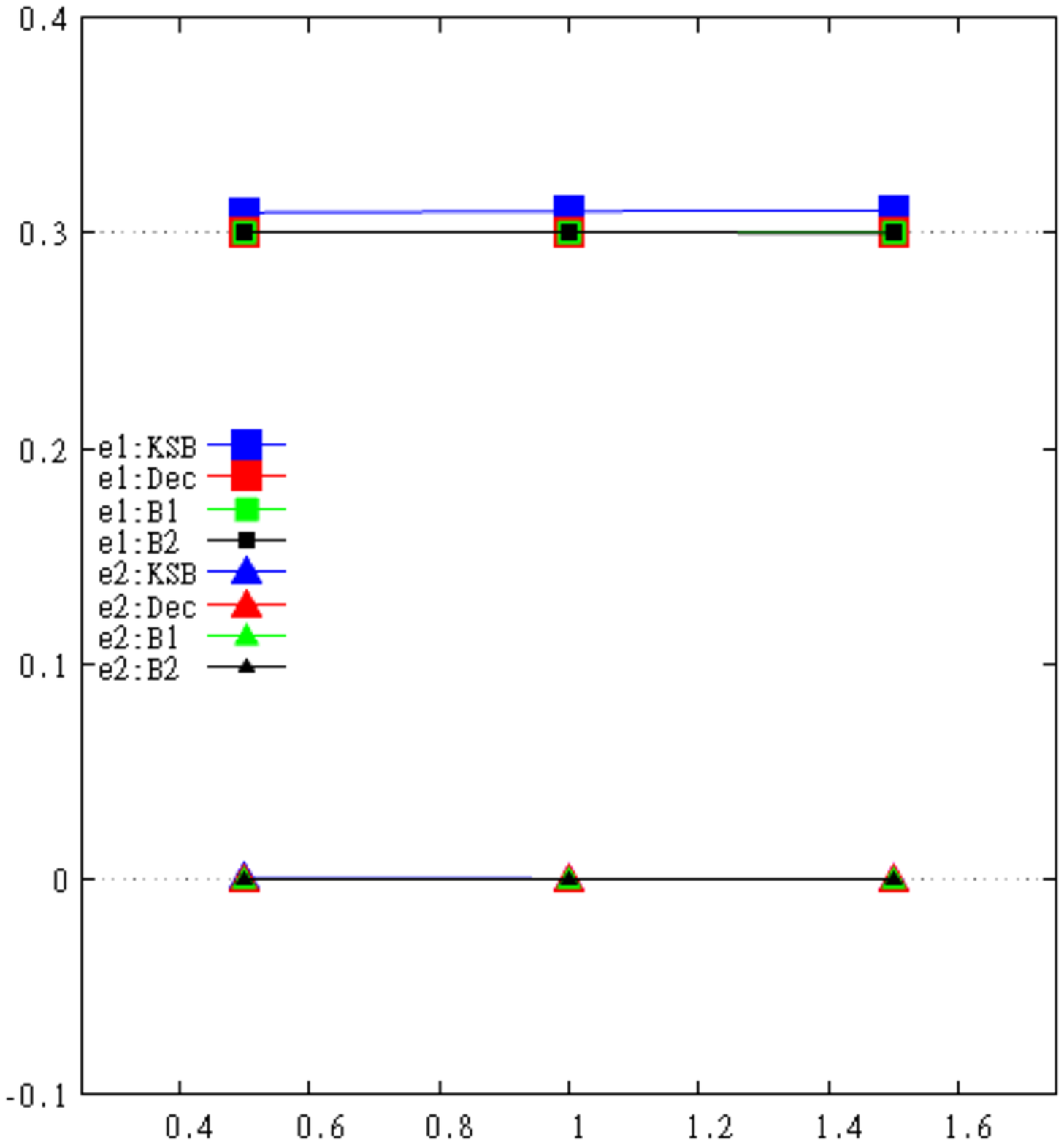}
\caption{
\label{fig:S_3_0A}
This figure is the same as Figure \ref{fig:S_0_0A} 
except the ellipticity of the PSF is (0.3, 0.0).
}
\end{figure*}

\begin{figure*}[htbp]
\epsscale{0.5}
\plotone{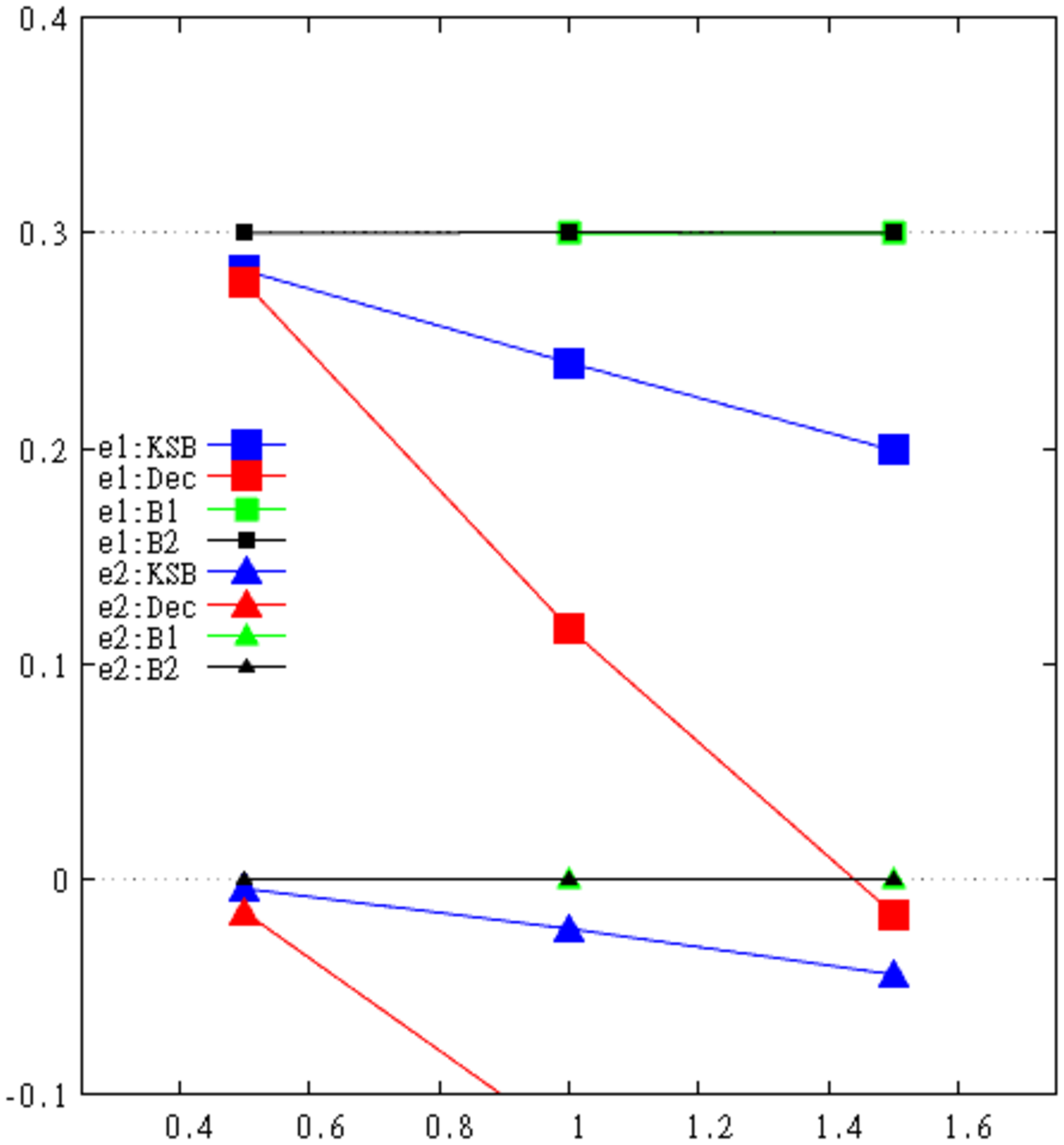}
\caption{
\label{fig:S_6_6A}
This figure is the same as Figure \ref{fig:S_0_0A} 
except the ellipticity of the PSF is (-0.6, -0.6).
}
\end{figure*}

\begin{figure*}[htbp]
\epsscale{0.5}
\plotone{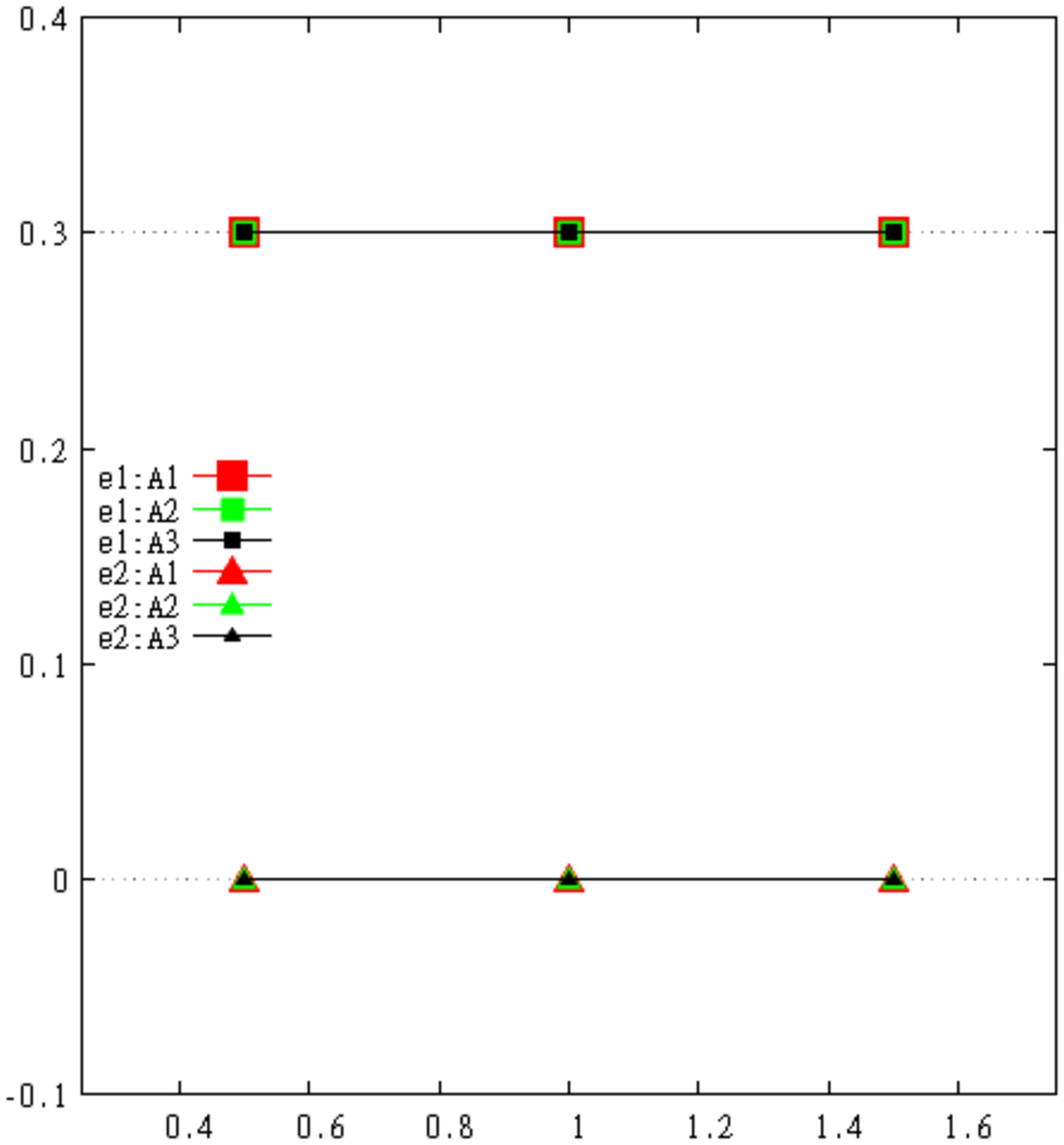}
\caption{
\label{fig:S_0_0B}
This figure is the same as figure \ref{fig:S_0_0A} 
except for the method used.
Red symbols (A1) mean PSF correction by Method A1,
green symbols (A2) mean by Method A2 and
black symbols (A3) mean by Method A3.
}
\end{figure*}

\begin{figure*}[htbp]
\epsscale{0.5}
\plotone{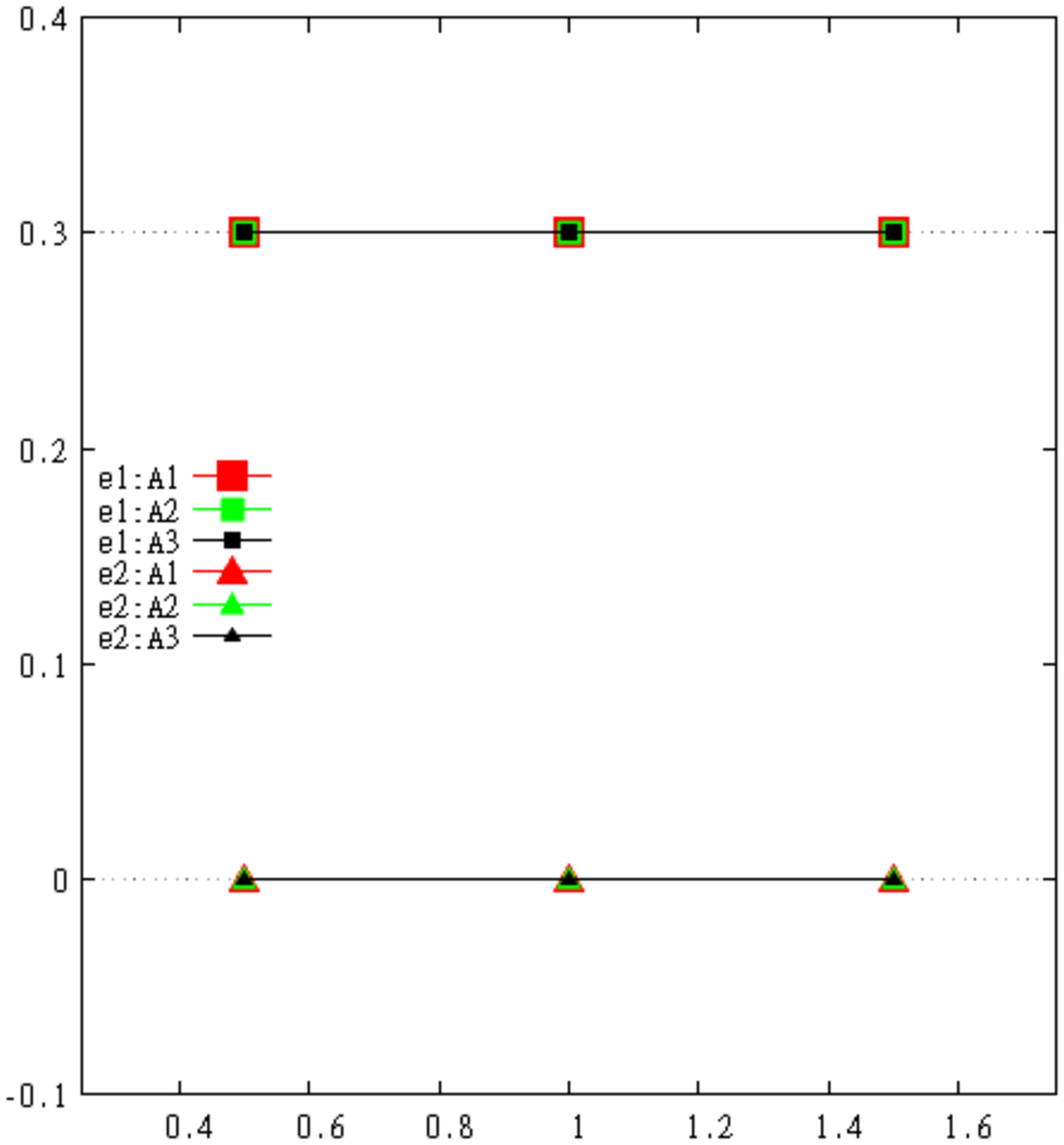}
\caption{
\label{fig:S_1_1B}
This figure is the same as Figure \ref{fig:S_0_0B} 
except the ellipticity of the PSF is (0.1, 0.1).
}
\end{figure*}

\begin{figure*}[htbp]
\epsscale{0.5}
\plotone{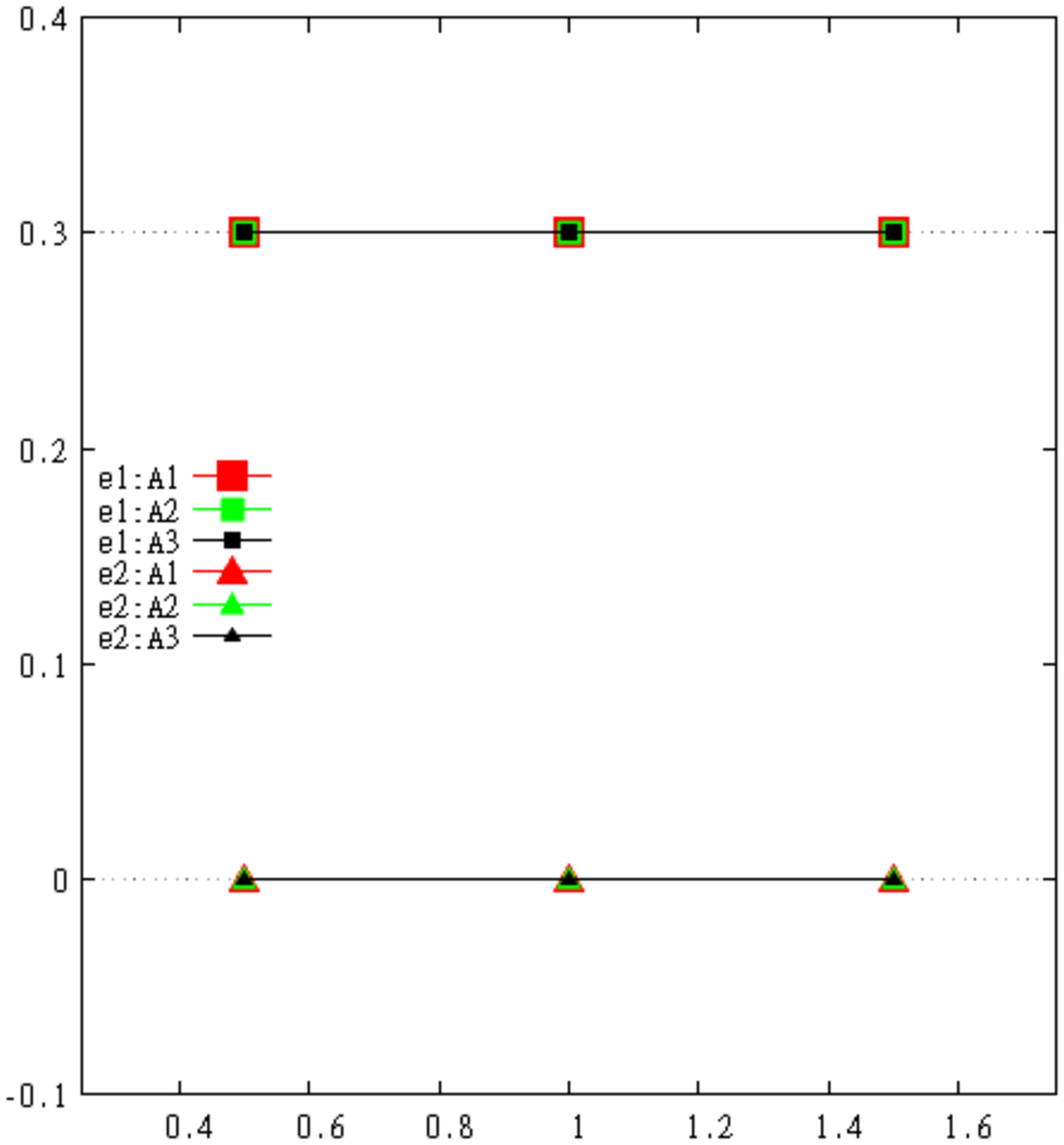}
\caption{
\label{fig:S_3_0B}
This figure is the same as Figure \ref{fig:S_0_0B} 
except the ellipticity of the PSF is (0.3, 0.0).
}
\end{figure*}

\begin{figure*}[htbp]
\epsscale{0.5}
\plotone{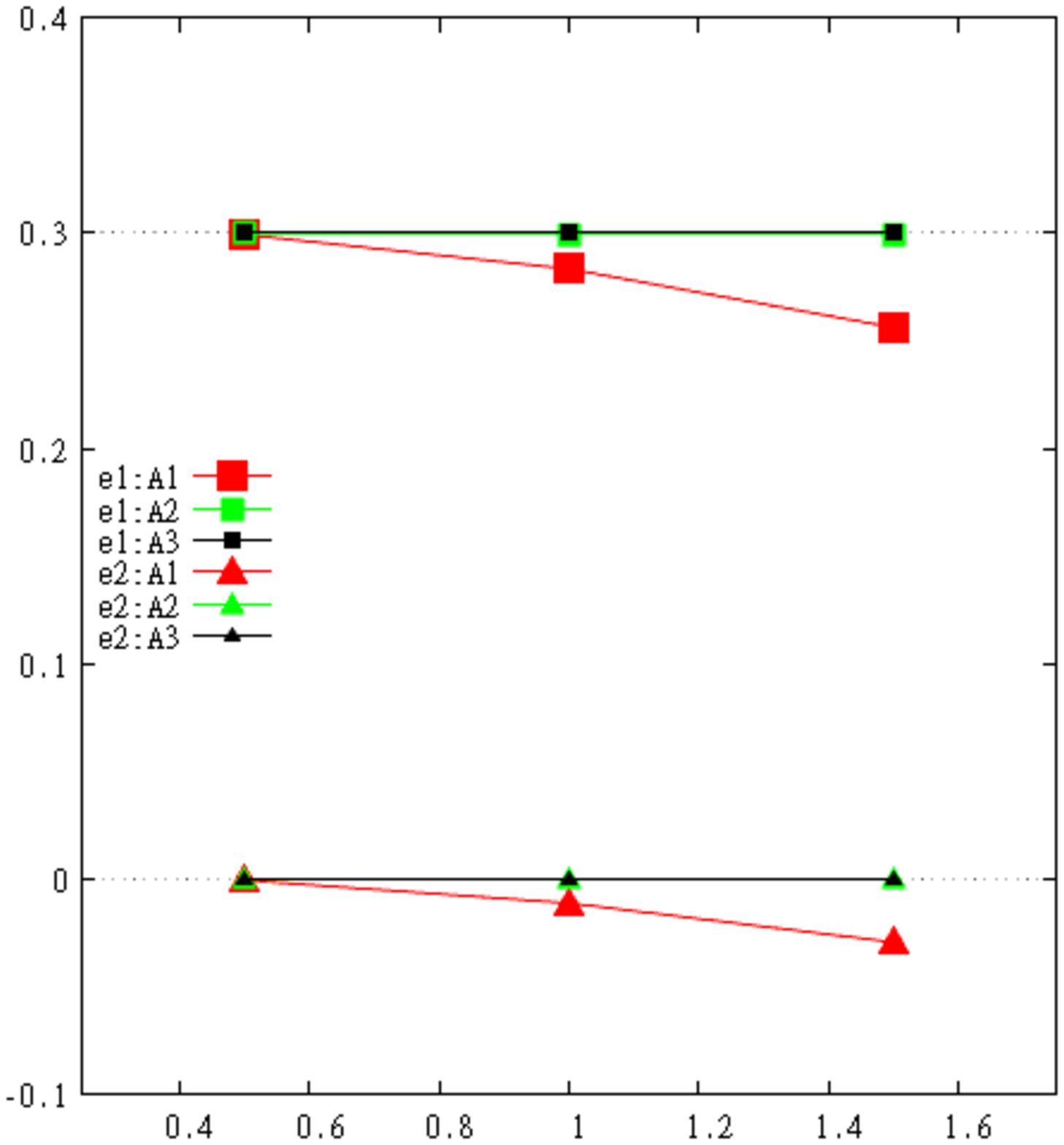}
\caption{
\label{fig:S_6_6B}
This figure is the same as Figure \ref{fig:S_0_0B} 
except the ellipticity of the PSF is (-0.6, -0.6).
}
\end{figure*}

Figures \ref{fig:G_0_0A} to  \ref{fig:S_6_6B} show the PSF corrected ellipticities.
The conditions of the lensed images and PSF are shown in each caption.
Fig.\ref{fig:G_0_0A}, Fig.\ref{fig:G_0_0B}, Fig.\ref{fig:S_0_0A} and Fig.\ref{fig:S_0_0B} show 
the results with an isotropic PSF, and
Fig.\ref{fig:G_1_1A}, Fig.\ref{fig:G_1_1B}, Fig.\ref{fig:S_1_1A} and Fig.\ref{fig:S_1_1B} show 
the results with a small elliptical PSF.

These results show that the standard deconvolution method has a systematic error in some situations, e.g. when there is a large PSF, 
and the KSB method overestimates in all situations.
In contrast, 
Method A and Method B correct the PSF with systematic errors of about 0.01\% or lower. Moreover,   
the error can be reduced with more iterations.
Fig.\ref{fig:G_3_0A}, Fig.\ref{fig:G_3_0B}, Fig.\ref{fig:S_3_0A} and Fig.\ref{fig:S_3_0B} show that 
all methods have no systematic errors,
because the ellipticity of the lensed images and that of the PSF have the same value,
so the PSF acts as $P^E$ and we do not need the PSF correction.
However, because KSB corrects both  anisotropic and isotropic PSF, the KSB method has a systematic error.
Fig.\ref{fig:G_6_6A}, Fig.\ref{fig:G_6_6B}, Fig.\ref{fig:S_6_6A} and Fig.\ref{fig:S_6_6B} show 
the results with a large elliptical PSF.
In real lensing analysis, such a large elliptical PSF is rare  or we should discard such data,
however, this sort of test is valuable to understand the details of this new method.
These results show that the 
deconvolution and KSB methods are not able to sufficiently correct  such a large ellipticity of the PSF. 
PSF corrections with a small re-smearing function in Method A have a systematic error.
Method A makes the re-smeared image from the deconvolved image, 
resulting in a large systematic error with the deconvolved method.
Method B1 with a PSF size 0.5 cannot correct the PSF because a high ellipticity (nearly 1) is 
needed for the re-smearing function $\Delta P$.
Method B2 has a larger  re-smearing function than Method B1, so Method B2 can correct the PSF with a 
smaller ellipticity for $\Delta P$ than Method B1. 
This indicates we have to choose the size of re-smearing function carefully 
when the PSF has  a large size and large ellipticity.
We obtained similar results using different ellipticity of the lensed image $\bep^L=$(0.1,0.0) and (0.5,0.0).
All figures show that Method A3 and Method B2 can correct the PSF  in all situations,
meaning that the ERA method with a large re-smearing function can correct the PSF. 
However we note that re-smearing by a re-smearing function that is too large, makes a large artificial image
and longer time is needed to measure the moments of this large artificial image.

However, it is important to find more appropriate profiles and sizes for $P^E$ to have  faster analyses,
and it is also necessary to find a more effective numerical iteration scheme. 
These problems will be studied in the future.


\section{Summary and Future Work}
We have developed the ERA method with a possible new PSF correction method for weak gravitational lensing shear analysis. 
The idea is to construct an artificial image with the same ellipticity as the lensed image by re-smearing 
the observed image. 
This approach avoids the approximations in PSF corrections of previous moments method (i.e. Eq.\ref{eq:KSBAP}),
therefore there is no systematic error from PSF correction if we choose an appropriate function for re-smearing.
Then, we tested the method with simple simulated images. 
The results of the simulation are as follows.
The deconvolution method cannot estimate ellipticity correctly, because deconvolution is not perfect when smearing using a large or high elliptical PSF.
The KSB method has systematic errors which cause  overestimation in the standard
situation (in the simulation, it is about 2-3\%, but it  depends on the 
ellipticity of the image and the size of the PSF, etc.), and is not able to correct the smearing effect with a high elliptical PSF.
However, 
Methods A and B are both able to estimate the ellipticity correctly if the PSF has standard ellipticity.
We also confirmed that re-smearing the deconvolved image (method A3) and re-smearing the smeared image (method B2) have 
no systematic error in PSF correction even if the size and the ellipticity of PSF are large.

Although the tests performed here are not entirely realistic, the results are very encouraging. 
Further studies of systematic errors based on realistic data, (e.g. pixel noise, pixelization etc.) are needed.
In particular, pixel noise has the potential to make large systematic errors (see E-HOLICs part3).
For example, in measuring cosmic shear,
galaxies in high redshift bins are relatively faint and have lower signal to noise ratio, and thus they 
suffer from more systematic bias due to pixel noise than those at lower redshifts. 
Thus, the measured shear for high-z galaxies would be underestimated. 
It is also important to find a more appropriate profile for the re-smearing function 
and to develop more effective iteration schemes which will result in faster analyses.  Future planed surveys 
such as EUCLID and LSST will treat an enormous number of galaxies and thus not only low systematic errors but also fast analyses 
are essential. These problems will be approached in future studies.

\acknowledgements

We would like thank Alan Lefor for improving English of this paper.


\end{document}